\documentclass[final,3p,times,twocolumn]{elsarticle}
\pdfoutput=1
\usepackage{subfigure}
\usepackage{array}
\usepackage{multicol}
\usepackage{lineno,hyperref}
\modulolinenumbers[5]

\usepackage{graphics}
\usepackage{graphicx}

\usepackage{amssymb}

\usepackage{tabularx}

\journal{Journal of Quantitative Spectroscopy \& Radiative
Transfer}

\begin{document}
\begin{frontmatter}
\title{Imaging polarimetry of the rotating Bok globule CB67}

\author[spbgu]{M.S.~Prokopjeva\corref{cor1}}
\ead{ari-76@yandex.ru}
\cortext[cor1]{Corresponding author.}

\author[iucaa,suniv]{A.K.~Sen}
\ead{asokesen@yahoo.com}

\author[spbgu,pulkovo,aerokos]{V.B.~Il'in}
\ead{ilin55@yandex.ru}

\author[spbgu]{N.V.~Voshchinnikov}
\ead{voshchinnikov@mpia-hd.mpg.de}

\author[iucaa]{R.~Gupta}
\ead{ranjaniucaa@gmail.com}

\address[spbgu]{St.~Petersburg State University, Astronomical Institute
Universitetskij pr. 28, St.Petersburg, 198504, Russia}

\address[iucaa]{Inter University Centre for Astronomy and Astrophysics, Ganeshkhind, Pune 411007, India}

\address[suniv]{Assam University, Silchar, 788011 India}

\address[pulkovo]{Main (Pulkovo) Astronomical Observatory, Pulkovskoe sh. 65/1,
St.Petersburg, 196140, Russia}

\address[aerokos]{St.~Petersburg State University of
Aerospace Instrumentation, ul. Bolshaya Morskaya 67,
St.Petersburg, 190000, Russia}

\date{\today}
\begin{abstract}

Polarimetric observations of about 50 stars located in a close
vicinity of the Bok globule CB67 having significantly nonspherical
shape and rapid rotation are performed. The data obtained are
compared with the available observations of this
globule at radio and submillimeter wavelengths as well as some
theoretical calculations. It is found that the elongation and the
rotation moment of CB67 are oriented rather perpendicular
to the magnetic fields, which is unusual for Bok globules
and is difficult to be explained from the
theoretical point of view.

\end{abstract}


\end{frontmatter}

\section{Introduction}
\label{sect1}

The Bok globules are known to be small molecular clouds where low
mass stars are formed. Despite a long interest in these rather
simple objects, details of the star formation process in them are
not well known (see, for example, \cite{Ber_2007, McK_2007,
Cer_2011}. Undoubtedly, the magnetic fields and to some extent
rotation should play an important role in the evolution of the
globules \cite{Hen_2013}.
 However, detailed observational data
are still required for more adequate understanding of the low mass
star formation.

Observations of polarization of background stars is the basic way
to study the magnetic fields in the vicinity and outer layers of
such clouds \cite{Cru_2012}. Though the first polarization maps of
Bok globules were obtained in the mid-eighties, only in a few
cases the data were sufficiently detailed (the number of stars was
about 30 or more). Like for more massive clouds the polarization
maps of globules have shown different behavior of the magnetic
fields (see for more details \cite{Ber_2007}). The connection of
the magnetic fields in the outer regions of globules with those in
their cores has been investigated by using the polarization maps
at submillimeter wavelengths \cite{Ward_2009}.

Information about angular velocity and the kind of rotation has
 also been obtained only for a few Bok globules (see, for example,
\cite{Kane_1997, Bel_2013}). The magnetic fields and rotation of
globules were observationally considered only by Kane and Clemens
\cite{Kane_1996}.
 Using polarization maps obtained for 6
globules with known rotation (the maps were not presented in the paper),
they concluded that the magnetic
fields, the rotation axis and the Galactic plane direction "tend
to be parallel". Note that for all
these globules the angle between the mean magnetic
fields and the rotation axis did not exceed 50$^\circ$.

In this paper a polarization map obtained by us for the
rapidly rotating Bok globule CB67 is presented.
 In Sect. \ref{sect2} and \ref{sect3} the basic
information about this cloud is given and the polarimetric data
derived for about 50 stars in its field are described.
 In Sect. \ref{sect4} we discuss the obtained results and point out that they
are untypical of globules.

\section{Object}
\label{sect2}

CB67 (L31) is a small isolated globule
(the center position: $l\approx1^\circ, b\approx+16^\circ$)
in the complex of molecular
clouds in Ophiuchus which is located at the distance of about
$\sim$120 pc \cite{Lom_2008}.
 The angular size of this globule on the POSS maps is
$16^\prime \times 4^\prime$,
the position angle
(hereafter all position angles are given in the equatorial coordinates)
of the large semiaxis equals about 110$^\circ$
\cite{Cle_1988}, the opacity class is 6 \cite{Lyn_1962}.

The globule has been observed in CO, $^{13}$CO, C$^{18}$O and OH
lines.
 These observations have shown that the globule velocity is about 4.7 km/s
with the velocity dispersions
$\sigma_{\rm v} ({\rm CO}) \approx 1.3$ km/s and
$\sigma_{\rm v} ({\rm^{13}CO}) \approx 0.9$ km/s.
 CB67 has the rapid differential rotation with the velocity gradient
$\nabla v \approx$ 2 km/s/pc,
the angular velocity $\omega = 7\, 10^{-14}$ s$^{-1}$  and the positional
angle of the rotation axis of
$\theta_{\rm J} = 112\pm1^\circ$ \cite{Kane_1997}.
 The parameters of two cores which were observed in $^{13}$CO line are similar: the size is about $14^\prime \times 5^\prime$, the density $n(H_2)
\approx$ 3\,$10^3$ cm$^{-3}$ and the mass M $\approx$ 7--8
M$_\odot$ \cite{Noz_1991}.

Infrared observation did not reveal protostellar sources in this
globule \cite{Lau_1997}. Visser et al.  have
investigated CB67 at submillimeter wavelengths and shown that the
emission region at 850 $\mu$m has the approximate size of
$11^\prime \times 1.5^\prime$ and mass of 2.2 M$_\odot$ \cite{Vis_2001}.
Note that these authors observed only a part (approximately a half) of the
globule, and the region of cold dust emission looks like a
filament with the typical diameter of about 0.06 pc
\cite{Arz_2011}.

So, the shape, size, location in the sky and other
characteristics of CB67 seem to be rather typical of Bok globules.

The large- and small-scale geometries of the Galactic magnetic
field are not yet well known \cite{Han_2013, Pavel_2012}.
 According to the standard stellar polarization data set
given by Heiles' catalog \cite{Heiles_2000}, the Galactic magnetic
field component $\vec{B}_\perp$ forms a large arc including the
points $(l,b)$ = $(300^\circ,0^\circ)$, $(330^\circ,-15^\circ)$,
$(360^\circ,0^\circ)$.
 So, in the Galactic plane
below CB67 ($l=0^\circ, b=-5-5^\circ$)
the mean position angle of stellar polarization
is about 170$^\circ$.

 The globule CB67 is close to the ring-shaped interface
between the Local Bubble and Loop I Bubble
at the distance 70--280 pc \cite{Choi_2013}.
 The polarization of stars in the area of this interface
was studied by Santos et al. \cite{Santos_2011}.
 They found that the polarization
position angle in their field A2 ($l = 355-15^\circ, b =
20-35^\circ$) varied from about 50 to 150$^\circ$ with the
preferable direction in a part of the interface ($l \approx
355-10^\circ, b \approx 17-20^\circ$) most close to CB67 being
characterized by $\theta \sim 65^\circ$.

\begin{figure}\center
\includegraphics[width=75mm]{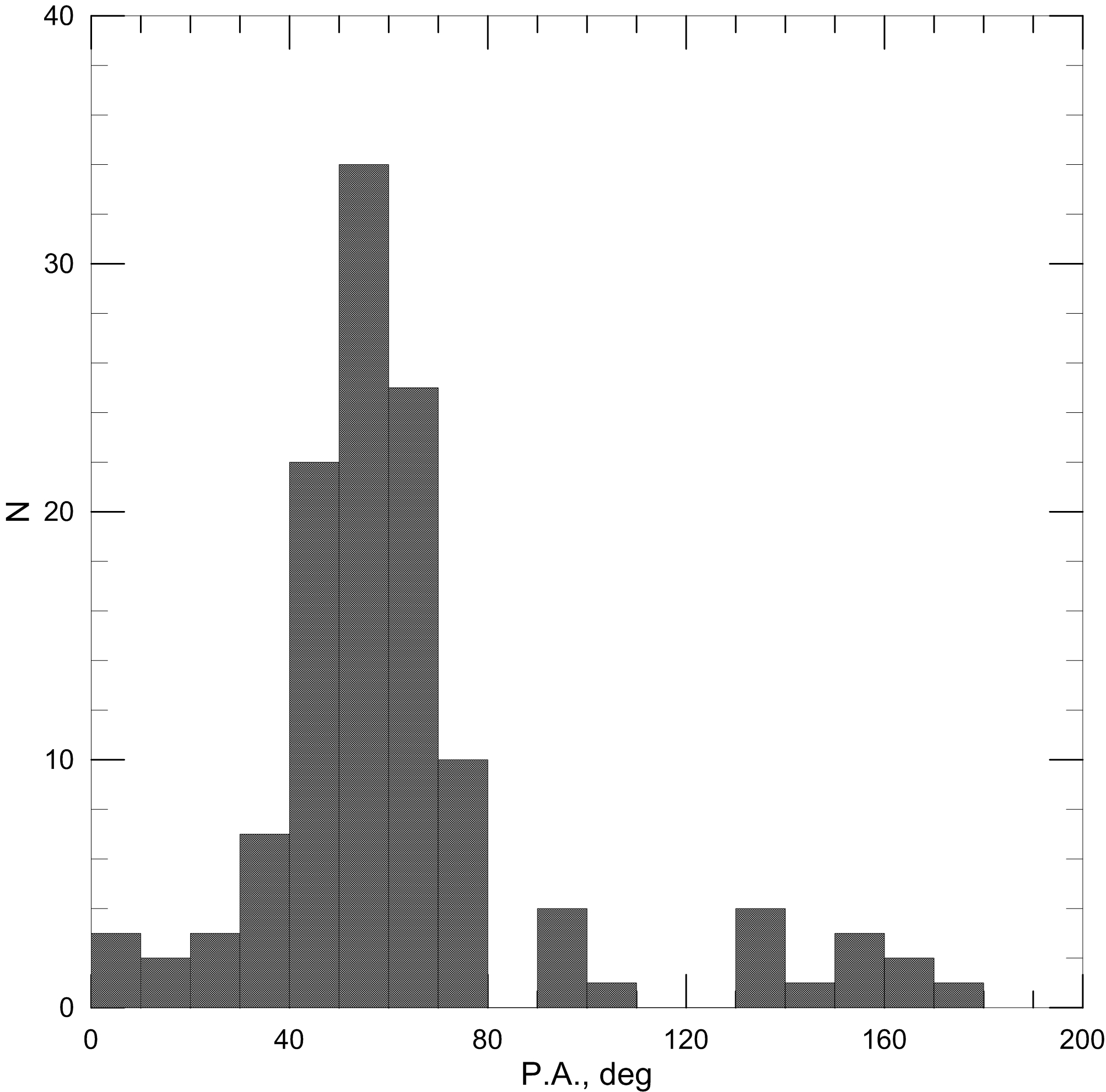}
\caption{Distribution of the position angle of polarization of
stars in a 7$^\circ$ radius circle around CB67
according to Heiles' catalog \cite{Heiles_2000}.
\label{Fig0}}
\end{figure}

The stellar polarization data given by Heiles \cite{Heiles_2000}
for the field $l = 350-10^\circ, b=7-27^\circ$ including the
Ophiuchus cloud complex were analyzed by Li et al. \cite{Li_2013}.
 They found that the mean polarization was
nearly parallel to the ``main filament''
with the position angle being equal to $62 \pm 26^\circ$,
which generally agrees with the polarization in the interface
region obtained in \cite{Santos_2011}.
 Using the same polarization data source, we have considered
the stellar polarization in a $7^\circ$ radius circle around CB67.
 The position angle distribution is given in Fig.~\ref{Fig0}.
The mean value equal to $60 \pm 15^\circ$ well agrees with
the results described above.

 Thus, we conclude that the positional angle of the mean Galactic
magnetic field component $\vec{B}_\perp$ in the Ophiuchus cloud complex
and in the Local Bubble -- Loop I Bubble interface
close to CB67 is about 60$^\circ$, which does not strongly differ
from the direction of the Galactic plane characterized
by $\theta_{\rm G} = 38.5^\circ$.

\section{Observations}
\label{sect3}

We performed polarimetric observations of stars in the vicinity of
CB67 using the 2-meter telescope of Girawali observatory (IUCAA)
in Pune (India) on March 12-14, 2013.
 The IUCAA Faint Object Spectrograph and Camera (IFOSC)
and the imaging polarimeter IMPOL \cite{Sen_1994,Ramaprakash_1998}
were used, the field of view had the diameter of 4$^\prime$, the
wavelength range was 0.35--0.8 $\mu$m.
 Some more details and references to the description of the camera work
in the polarization mode can be found in  \cite{Pau_2012}.

For the polarimetric standards HD94851 and HD43384, we
performed observation in the B and V bands and
obtained the following values of the polarization degree:
$P_{\rm B}=0.065\pm0.05$\% for the first star and
$P_{\rm V}=2.936\pm0.019$\% (with the position angle error
of 0.2$^\circ$) for the second star.
 These values well agree with data from the literature:
$P_{\rm B}=0.057 \pm 0.18$\%
for HD94851 \cite{Tur_1990} and $P_{\rm V}=2.94\pm0.04$\%,
$\theta_V = 169.8 \pm 0.7^\circ$ for HD43384 \cite{Hsu_1982}.
 The position angle of the latter standard was used to calibrate
the position angles of polarization observed.

\begin{figure}\center
\includegraphics[width=80mm]{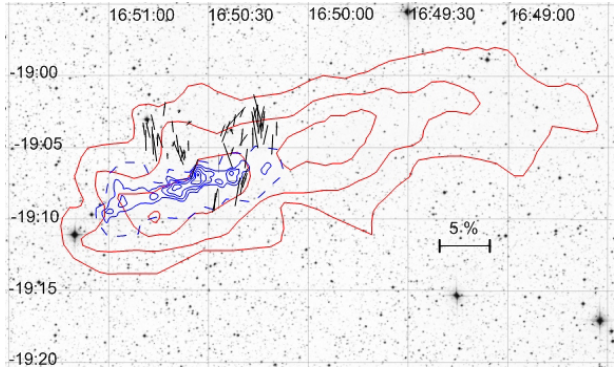}
\caption{Polarization of stars in the region of CB67.
  The dotted lines show emission in the $^{13}$CO line
\cite{Noz_1991}, the solid lines emission at $\lambda = 850$
$\mu$m \cite{Vis_2001}.
 The dashed line confines the field
observed in the submillimeter range. \label{Fig1}}
\end{figure}

The instrumental polarization of the IFOSC on the 2-m Giravali
telescope has been monitored for many years and is known to be
less than 0.05\% (see, e.g., \cite{Eswaraiah_2013}), which is also
confirmed by our results for the unpolarized standard.
 As the instrumental polarization
is such small and does not behave like a systematic error, it is not
being subtracted. Note also that our conclusions below
are based on data with $P > 1$\% when uncertainty of the position angle
caused by the instrumental polarization should be below 1.5$^\circ$.


Our polarimetric observations of three fields of 4$^\prime$
diameter were made without the use of any filters.
 Data were processed in the standard way.
 The results are presented in
Fig.~\ref{Fig1} where the polarization vectors for 49 stars (the
vector length is proportional to the polarization degree) and
contours of intensity in the $^{13}$CO line \cite{Noz_1991} and at
850 micron \cite{Vis_2001} are shown. Table~\ref{Table1} contains
the coordinates, polarization degree and position angle for
the stars shown in Fig.~\ref{Fig1}.

\begin{figure}[t]
\center
\includegraphics[width=80mm]{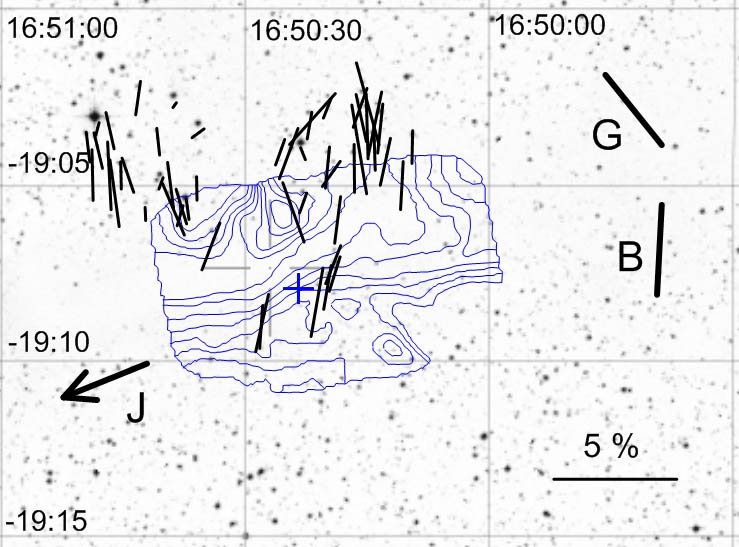}
\caption{Rotation of CB67 according to \cite{Kane_1997}.
 The solid lines are contours of the mean radial velocity.
 Location of the center of motion
(the peak of $^{13}$CO integrated intensity) is given by the
cross.
 The projected angular momentum $\vec{J}$,
the Galactic plane direction $\vec{G}$, 
and the average magnetic field direction $\vec{B}_\perp$ are
presented by the corresponding vectors.
 Polarization of the stars observed is also given.
\label{Fig2}}
\end{figure}

Note that some of our stars are located in the region of CB67 that
was considered by Kane and Clemens  when they studied its rotation
\cite{Kane_1997}.
 In Fig.~\ref{Fig2} we present the figure from their paper after adding the
vector of rotation moment, the line parallel to the Galactic
plane, the polarization vectors for stars observed by us and the
average direction of the magnetic fields near CB67 according to
our results.

\begin{table}[p!]
\caption{Polarization measurements of stars near CB67}
\newcolumntype{Y}{>{\footnotesize}r}
\newcolumntype{Z}{>{\footnotesize}c}
\setlength{\tabcolsep}{3pt}
\renewcommand{\arraystretch}{0.96}
\begin{tabularx}{80mm}[t]{>{\setlength{\hsize}{.3\hsize}}Y>{\setlength{\hsize}{1.0\hsize}}Z>{\setlength{\hsize}{1.0\hsize}}Z>{\setlength{\hsize}{.5\hsize}}Y>{\setlength{\hsize}{.5\hsize}}Y>{\setlength{\hsize}{0.8\hsize}}Y>{\setlength{\hsize}{.6\hsize}}Y}
\cline{1-7}\\
 No. & $\alpha$, h & $\delta$, deg & $P,\%$ & ${\rm err}_P$, \% & $\theta$, deg & ${\rm err}_\theta$, deg
 \\ \\
\cline{1-7}
\\
 1 & 16.840147 & -19.061911 & 0.393 & 0.041 & 201.823 &  2.989 \\
 2 & 16.837495 & -19.046930 & 2.111 & 0.074 & 174.043 &  1.010 \\
 3 & 16.837724 & -19.056358 & 1.908 & 0.096 & 181.931 &  1.434 \\
 4 & 16.838400 & -19.065884 & 2.039 & 0.180 & 180.529 &  2.526 \\
 5 & 16.839283 & -19.066020 & 1.737 & 0.254 & 174.395 &  4.186 \\
 6 & 16.837975 & -19.044588 & 2.844 & 0.242 & 181.629 &  2.441 \\
 7 & 16.837627 & -19.043412 & 2.833 & 0.458 & 168.032 &  4.630 \\
 8 & 16.839890 & -19.047447 & 1.065 & 0.305 & 170.202 &  8.194 \\
 9 & 16.840097 & -19.084549 & 0.782 & 0.246 & 160.286 &  9.002 \\
13 & 16.836435 & -19.057038 & 1.330 & 0.540 & 178.830 & 11.641 \\
15 & 16.839278 & -19.039856 & 0.772 & 0.373 & 158.423 & 13.835 \\
16 & 16.838130 & -19.037979 & 1.726 & 0.500 & 160.545 &  8.301 \\
17 & 16.837295 & -19.062799 & 2.050 & 0.254 & 167.744 &  3.551 \\
18 & 16.836777 & -19.073952 & 2.054 & 0.814 & 176.928 & 11.360 \\
19 & 16.839089 & -19.070290 & 1.237 & 0.550 & 144.600 & 12.729 \\
21 & 16.838930 & -19.090313 & 1.996 & 0.587 & 172.756 &  8.426 \\
22 & 16.840571 & -19.085646 & 2.761 & 0.471 & 199.778 &  4.886 \\
23 & 16.840515 & -19.058644 & 2.312 & 0.895 & 159.596 & 11.085 \\
24 & 16.840861 & -19.060060 & 1.067 & 0.784 & 160.231 & 21.040 \\
25 & 16.839561 & -19.042167 & 3.009 & 0.972 & 138.034 &  9.253 \\
26 & 16.838070 & -19.032501 & 3.805 & 1.237 & 195.397 &  9.315 \\
27 & 16.838316 & -19.053435 & 4.020 & 2.120 & 189.485 & 15.108 \\
28 & 16.838219 & -19.042146 & 2.060 & 1.077 & 192.321 & 14.973 \\
35 & 16.845459 & -19.090318 & 0.403 & 0.023 & 182.582 &  1.600 \\
36 & 16.847414 & -19.057360 & 1.194 & 0.066 & 186.862 &  1.579 \\
37 & 16.847092 & -19.058471 & 1.511 & 0.066 & 193.316 &  1.250 \\
38 & 16.846283 & -19.070901 & 1.071 & 0.048 & 181.209 &  1.295 \\
39 & 16.846108 & -19.070207 & 1.493 & 0.095 & 198.818 &  1.821 \\
40 & 16.846694 & -19.049021 & 1.539 & 0.199 & 192.637 &  3.703 \\
41 & 16.846694 & -19.067703 & 2.977 & 0.382 & 182.964 &  3.677 \\
42 & 16.843722 & -19.077402 & 0.886 & 0.128 & 181.049 &  4.149 \\
43 & 16.844296 & -19.084957 & 1.497 & 0.320 & 196.845 &  6.119 \\
44 & 16.844259 & -19.090532 & 0.765 & 0.312 & 163.644 & 11.688 \\
46 & 16.845681 & -19.033881 & 1.378 & 0.563 & 171.910 & 11.707 \\
47 & 16.846549 & -19.076002 & 2.952 & 0.450 & 187.862 &  4.364 \\
48 & 16.847249 & -19.068779 & 2.174 & 0.591 & 181.334 &  7.790 \\
49 & 16.844628 & -19.084826 & 2.424 & 0.710 & 202.968 &  8.387 \\
50 & 16.844639 & -19.073954 & 1.943 & 0.950 & 187.438 & 14.004 \\
51 & 16.845038 & -19.055047 & 1.050 & 0.761 & 186.038 & 20.759 \\
52 & 16.839006 & -19.116860 & 1.491 & 0.632 & 162.659 & 12.136 \\
53 & 16.839593 & -19.127341 & 3.056 & 0.551 & 170.128 &  5.161 \\
54 & 16.839267 & -19.122442 & 1.876 & 0.787 & 170.271 & 12.016 \\
55 & 16.839034 & -19.112131 & 1.711 & 0.563 & 158.491 &  9.422 \\
56 & 16.841444 & -19.138138 & 2.555 & 0.691 & 167.358 &  7.753 \\
57 & 16.841530 & -19.141177 & 1.813 & 0.836 & 175.973 & 13.211 \\
58 & 16.843146 & -19.102738 & 2.157 & 0.811 & 159.042 & 10.765
\\\\
\cline{1-7}
\end{tabularx}
\label{Table1}
\end{table}

Using $JHK$ data available for some of our stars in the Two-Micron
All-Sky Survey catalog \cite{Skrutskie_2006}, we have roughly
estimated the distances $d$ and visual extinction $A_{\rm V}$ for
about 20 stars observed (see Table~\ref{Table2}) following the
approach developed in \cite{Maheswar_2010}.
 For other our stars, either $JHK$ data were absent, or
$(J-K) > 0.75$, which makes such spectral class
estimates less reliable.
 Two stars (N 29 and 32) were not included in Table~\ref{Table1}
and our figures shown above because of their large polarization
errors:
 $P = 0.4 \pm 1.8$\% for N 29 and
 $P = 0.8 \pm 1.3$\% for N 32.
 The values of $A_{\rm V}$ were also estimated utilizing
the extinction map obtained from 2MASS data
with the NICE technique in \cite{Rowles_2009}.
 A weak correlation of two extinction estimates
given in Table~\ref{Table2} can be seen.

\begin{table}[tb]
\caption{Estimates of distance and visual extinction from $JHK$
photometry}
\newcolumntype{Y}{>{\footnotesize}r}
\newcolumntype{Z}{>{\footnotesize}c}
\newcolumntype{W}{>{\footnotesize}l}
\setlength{\tabcolsep}{3pt}
\renewcommand{\arraystretch}{0.96}
\begin{tabularx}{70mm}[t]{>{\setlength{\hsize}{.3\hsize}}Y>{\setlength{\hsize}{1.0\hsize}}Z>{\setlength{\hsize}{1.0\hsize}}Z>{\setlength{\hsize}{.5\hsize}}Y>{\setlength{\hsize}{.5\hsize}}Y>{\setlength{\hsize}{0.8\hsize}}Y>{\setlength{\hsize}{.6\hsize}}Z>{\setlength{\hsize}{.6\hsize}}W}
\cline{1-8} \\
 No. & $P, \%$  & $J$ & $H$ & $K$ & $d$, pc & $A_{\rm V}$, mag & $A_{\rm V}$, mag \cite{Rowles_2009} \\
\\
\cline{1-8}\\
 1 & 0.39\% & 10.416 & 10.096 & 9.901  &  400 & 2.18 & 1.71 \\
 7 & 2.83\% & 14.343 & 13.827 & 13.608 & 1010 & 2.14 & 1.61 \\
 9 & 0.78\% & 13.148 & 12.805 & 12.632 & 1030 & 1.65 & 1.87 \\
15 & 0.77\% & 13.855 & 13.285 & 13.129 &  560 & 0.41 & 1.55 \\
23 & 2.31\% & 14.775 & 14.308 & 14.091 & 1560 & 2.25 & 1.61 \\
27 & 4.02\% & 15.549 & 15.137 & 15.002 & 1600 & 0.69 & 1.63 \\
29 & 0.39\% &  8.649 &  8.346 &  8.251 &   95 & 0.18 & 0.81 \\
32 & 0.77\% & 10.680 & 10.327 & 10.152 &  360 & 1.67 & 1.46 \\
37 & 1.51\% & 13.235 & 12.690 & 12.530 &  460 & 0.69 & 0.92 \\
40 & 1.54\% & 14.063 & 13.597 & 13.473 &  730 & 0.18 & 0.81 \\
41 & 2.98\% & 14.691 & 14.236 & 14.021 & 1600 & 2.24 & 0.97 \\
43 & 1.50\% & 15.227 & 14.744 & 14.486 & 2260 & 3.01 & 1.46 \\
48 & 2.17\% & 14.407 & 14.079 & 13.943 & 1650 & 0.94 & 0.95 \\
49 & 2.42\% & 15.349 & 15.031 & 14.901 & 2730 & 0.89 & 1.40 \\
50 & 1.94\% & 15.32  & 14.864 & 14.684 & 1630 & 1.55 & 1.33 \\
51 & 1.05\% & 15.177 & 14.603 & 14.441 & 1010 & 0.54 & 1.14 \\
53 & 3.06\% & 14.287 & 13.747 & 13.541 &  770 & 1.80 & 1.55 \\
57 & 1.81\% & 14.328 & 13.857 & 13.681 &  930 & 1.35 & 1.79 \\\\
\cline{1-8}
\end{tabularx}
\label{Table2}
\end{table}

Using Table~\ref{Table2} we produced Fig.\ref{Fig3} that shows the
dependences of $P$ and $P/A_{\rm V}$ on $A_{\rm V}$.
 Excluding the stars N 27 and 40 for which we probably derive
too low extinction $A_{\rm V} = 0.18$, our Fig.\ref{Fig3}
demonstrates the trends typical of other globules (see, e.g.,
\cite{McCutcheon_1986, Eswaraiah_2013}): 1) a growth of maximum of
the polarization degree $P$ (in percent) with $A_{\rm V}$ (in
mag.) so that $P < 2 A_{\rm V}$; 2)~a decrease of maximum of the
polarization efficiency $P/A_{\rm V}$ with $A_{\rm V}$ so that
$P/A_{\rm V} < -1.5 A_{\rm V} + 5$.
 Note that as the wavelength $\lambda_{\rm max}$
of maximum polarization $P_{\rm max}$
for stars such close to globules usually
is in the interval about 0.5--0.65 $\mu$m (e.g., \cite{Eswaraiah_2013})
and for the standard Serkowski law of the wavelength dependence
of interstellar polarization in the interval
$\lambda/\lambda_{\rm max} = 0.75 - 1.3$ the values of $P(\lambda)$
differ from $P_{\rm max}$ by less than 10\%,
we probably have $P \approx P_{\rm max} \approx P_{\rm V}$
where $P$ is the polarization degree derived by us without a filter
in the instrument range 0.35--0.80 $\mu$m.
 Note also that according to our estimates all stars in Table~\ref{Table2}
except for N 29 are located at distances larger than $\sim 400$ pc
as expected from a statistical point of view.

\section{Discussion}
\label{sect4}

Figure~\ref{Fig2} shows that the vectors of observed polarization
are oriented rather uniformly in the field of CB67. It is better
seen in Figs.~\ref{Fig4}--\ref{Fig5} where we present some
distributions over the position angle. Note that the polarization
degree is mainly within the range 0.8--3\% and the position angle
$\theta$ in the interval
 150--200$^\circ$,
with the average position angle being $\bar{\theta} =
176.7^\circ$ and the dispersion $\sigma_\theta = 14.8^\circ$.

 The distribution of stars over $\theta$ is rather typical of globules
 (see, for example, a similar distribution for globule B227
 in \cite{Eswaraiah_2013}).
Note, however, that there are many globules where the distribution
of polarization vectors is not such uniform \cite{Kane_1996}.

The important question is where the polarization we observe is
originated? Generally, stellar polarization can provide information
about the magnetic fields in dusty media located foreground or
background of a cloud complex or within the complex but not
physically related to the cloud studied.

A study of stellar polarization performed in \cite{Santos_2011}
has shown that for their field I2 corresponding to the
Ophiuchus molecular cloud complex
for all (about couple of dozens) stars
at the distance $d < 100$ pc one has the polarization
degree $P < 0.05$\%, and
for stars at $d > 120$ pc \ $P$ is in the interval 0--2\%.
 As for all (except for two) stars observed we got $P > 0.8$\%
we can conclude that the contribution of the foreground material
to observed polarization should be negligible in the case of CB67.

 The foreground polarization can be estimated for more directions
from a relation between the interstellar extinction and polarization.
 But both the estimates based on Str\"omgren photometry and
Hipparcos parallaxes in \cite{Vergely_2010} and those derived from
interstellar lines of Na I and Ca II in \cite{Welsh_2010}
do not
show any significant extinction in the direction of $l \approx
0^\circ$ up to about 100 pc where the Ophiuchus cloud complex
appears.

\begin{figure*}
\center
\subfigure[]{\includegraphics[width=0.47\textwidth]{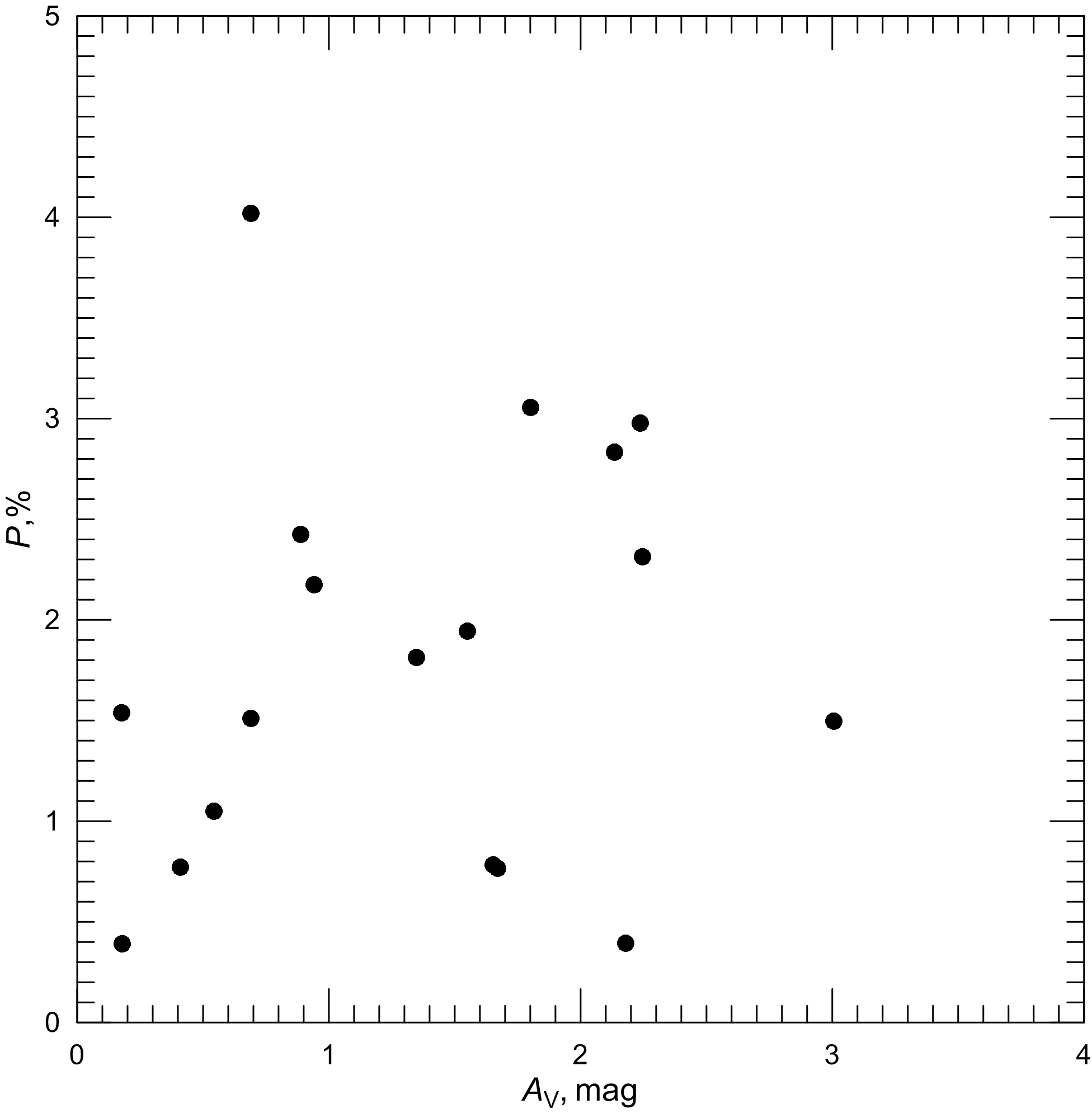}}\hfill
\subfigure[]{\includegraphics[width=0.47\textwidth]{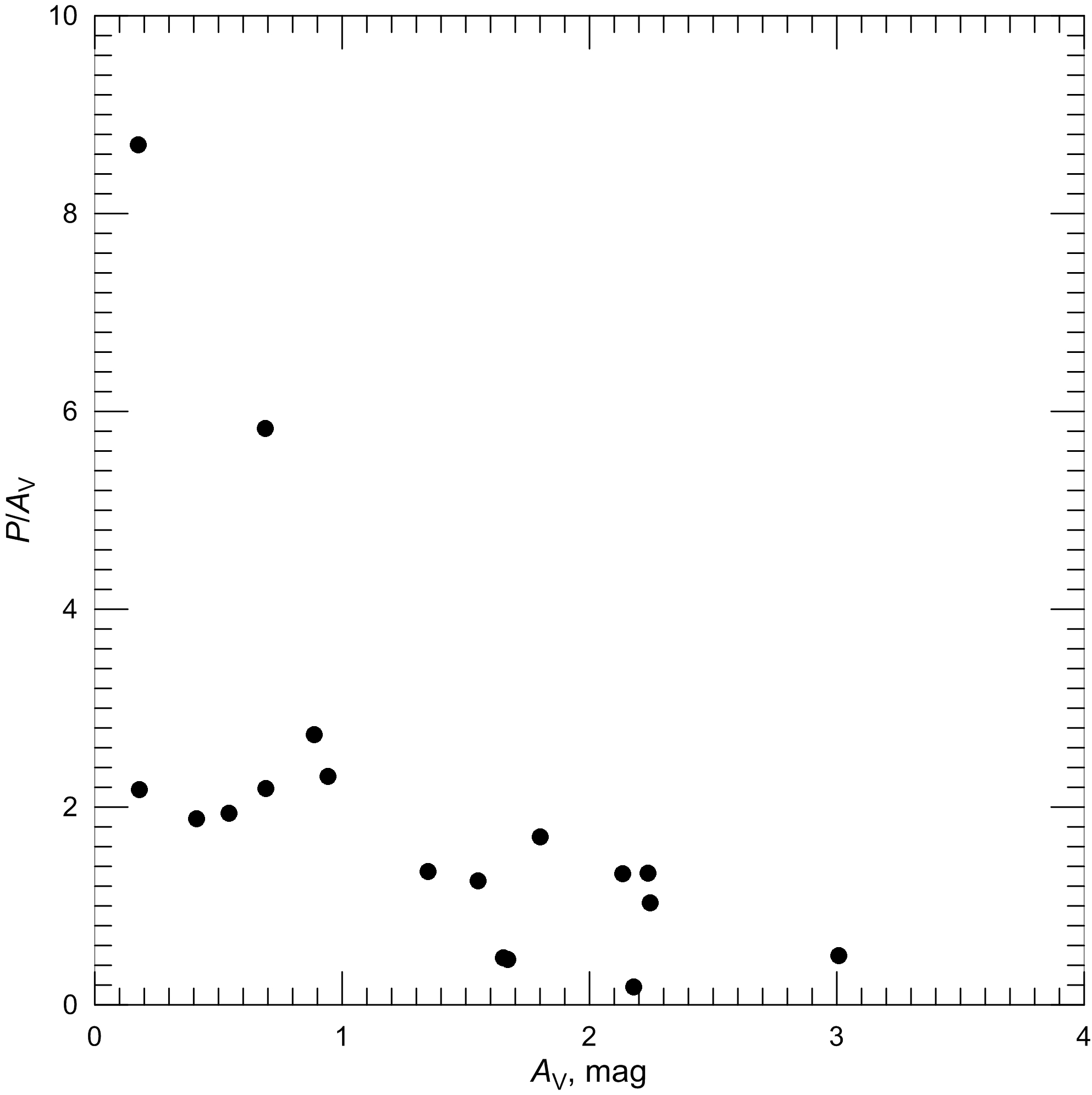}}
\caption {Dependence of polarization degree $P$ on visual
extinction $A_{\rm V}$ (a) and dependence of polarization
efficiency $P/A_{\rm V}$ on $A_{\rm V}$ (b) for some stars near
CB67} \label{Fig3}
\end{figure*}



\begin{figure*}
\includegraphics[width=0.47\textwidth]{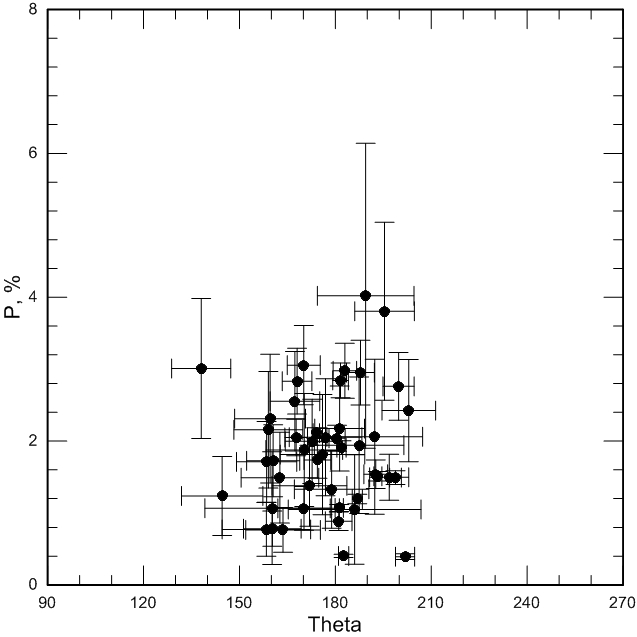}
\hfill
\includegraphics[width=0.47\textwidth]{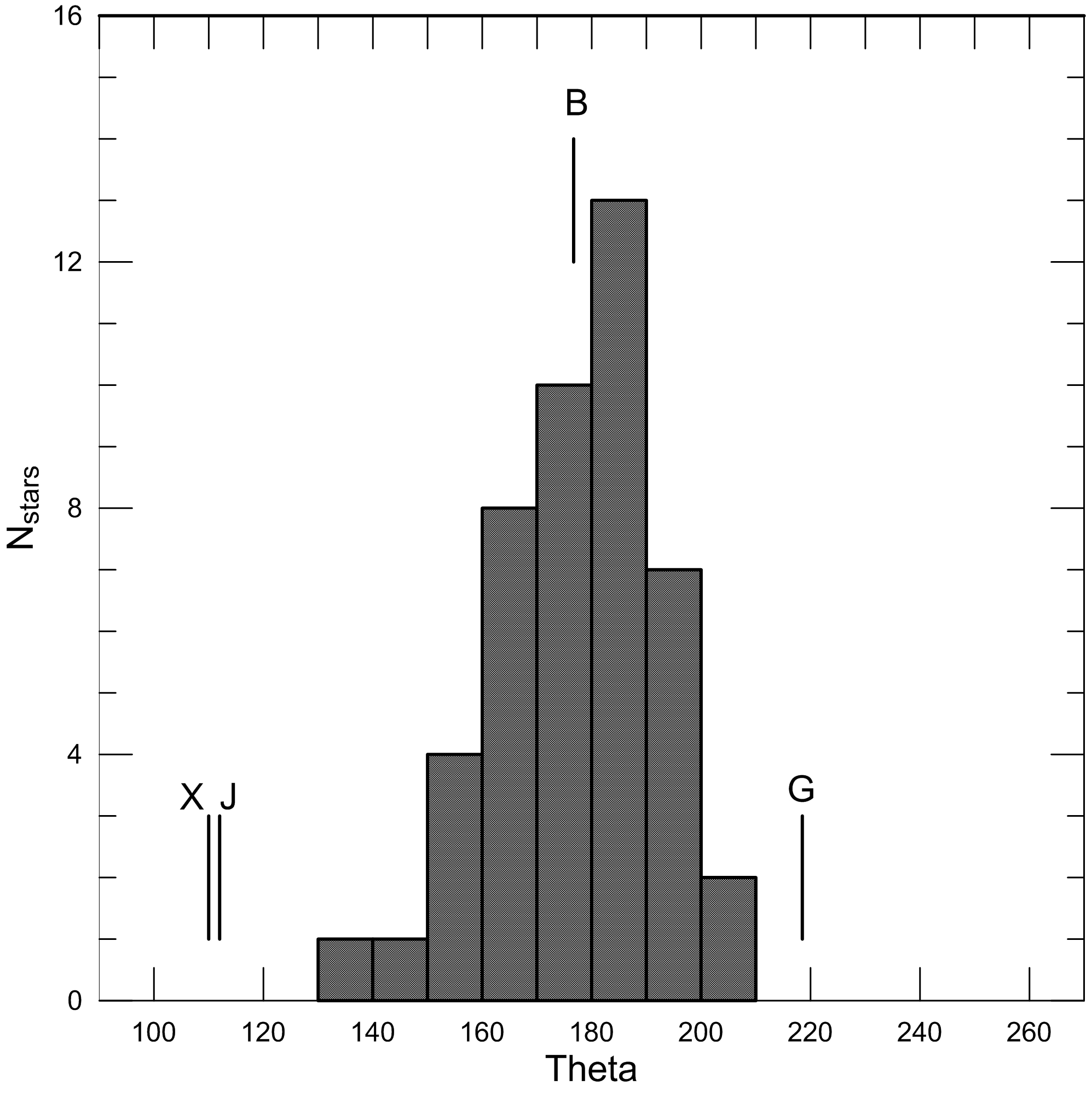}
\parbox[t]{75mm}{ \caption{Polarization degree $P$
in dependence on the positional angle $\theta$ for all stars
observed.}\label{Fig4}} \hfill
\parbox[t]{75mm}{
\caption{ The distribution of all stars observed over $\theta$.
The labels show the position angles of the largest extension in
visual (X), the angular moment (J), the Galactic plane (G), and
the mean magnetic field (B).} \label{Fig5}}
\end{figure*}


 The contribution of background polarization hardly can be well estimated.
 The line of sight to CB67 ($l \approx 1^\circ, b \approx 16^\circ$)
is directed generally above the Galactic center.
 Looking in the Galactic plane we see
the Carina-Sagittarius arm at $d$ about 0.7--1.4 kpc and the
Crux-Scutum arm at $d$ about 3 kpc.
 As Table~\ref{Table2} predicts the stars observed should mainly belong
to the interarm space and the Carina-Sagittarius arm.
 As the CB67 latitude is high enough we can agree with
\cite{Li_2013} 
that there should not be much diffuse dust outside
the Ophiuchus molecular cloud complex
to produce essential interstellar (extinction and) polarization.
 Additionally, all dark clouds with the estimated distance in
the field of the Ophiuchus complex have $d \sim 150$ pc
(see, e.g., \cite{Hilton_1995} and references therein), and
there are no signs that another more distant dense cloud is
projected on the CB67 region.

Considering emission in the $^{13}$CO line studied in
\cite{Noz_1991}, we see that the stars observed are projected well
inside the contours related with CB67 (see Fig.~\ref{Fig1}).
 Additionally, about 1/3 of our stars is projected on the contours
of the systematic motion (rotation) of CB67 observed
in \cite{Kane_1997}. 
 Obviously, this denser gas and the dust mainly responsible
for the observed polarization should spatially coincide.
 A weaker argument for the relation of the polarization with CB67
is that the dependences $P$ vs. $A_{\rm V}$ and $P/A_{\rm V}$ vs.
$A_{\rm V}$ presented in Fig.~\ref{Fig3} well resemble those
obtained for similar dense clouds.

Thus, we assume that the mean direction of observed polarization
characterizes the magnetic field in the vicinity of CB67,
and its position angle $\theta_{\rm B}$ is equal to
$\bar{\theta}$ obtained above.
 Then it is a bit unexpected that
from our analysis in Sect.~2 it follows that
the direction of the local field component $\vec{B}_\perp$ near CB67
($\theta_{\rm B} = 177^\circ$)
is parallel to the global field in the Galactic plane
($\theta \approx 170^\circ$)
and nearly perpendicular to the magnetic field
in the Local Bubble and
Loop I Bubble interface ($\theta \approx 60-65^\circ$)
being very close spatially to CB67.


Now it is interesting to compare 4 directions in the case of CB67:
those of the average magnetic fields
in the close vicinity of the globule
(according to our results
its position angle is          $\theta_{\rm B} = 177 \pm 15^\circ$),
the rotational angular momentum
                              ($\theta_{\rm J} = 112 \pm 1^\circ$ \cite{Kane_1997}),
the Galactic plane            ($\theta_{\rm G} = 38^\circ$)
and the elongation of the globule image
on the $^{13}$CO maps \cite{Noz_1991}
                              ($\theta_{\rm x}^{\rm CO} = 110 \pm 7^\circ$).
The directions of the largest extension in visual
and at submillimeter wavelengths have similar values:
$\theta_{\rm x}^{\rm vis} = 110^\circ$ and
$\theta_{\rm x}^{\rm submm} = 104^\circ$, respectively.
 Note that according to Li et al. (2013)
the direction of ``the main filament'' of the Ophiuchus
molecular cloud complex has $\theta_{\rm fil} = 70 \pm 12^\circ$.

For CB67, the difference $|\theta_{\rm B} - \theta_{\rm x}| =
67 \pm 13^\circ$ is rather close to 90$^\circ$. For other extended
molecular clouds, this difference is close to either 0 or 90$^\circ$
 \cite{Ber_2007}.
 Li et al. \cite{Li_2013} have considered 12 cloud complexes,
entering into the Gould belt,
 and found that the main filament direction and the average
magnetic field one differ by either less than 10$^\circ$ (4
complexes), or more than 70$^\circ$ (7 complexes, and in 5 cases
this difference exceeds 85$^\circ$).
 On other side, Ward-Thompson et al. \cite{Ward_2009} have noted that
 in the cores of low mass molecular clouds (7 ones were considered)
the difference of the small semiaxis direction and the magnetic
field one obtained from submillimetric polarimetry data is
about 30$^\circ$ with a small dispersion.
 If the field inside CB67 were parallel to that
in its vicinity, we would have $|\theta_{\rm B}^{\rm submm} -
\theta_{\rm y}^{\rm submm}| \approx 15^\circ$.

We get $|\theta_{\rm J} - \theta_{\rm G}| \approx 30 \pm 1^\circ$.
Note that Kane and Clemens \cite{Kane_1997} have considered 14
globules and found that this difference takes values from 0 to
90$^\circ$ with approximately equal probability, though the values
in the range 40--50$^\circ$ may be a bit more probable.

The difference $|\theta_{\rm B} - \theta_{\rm J}| = 65 \pm
15^\circ$ is remarkable.
 So far only Kane and Clemens have systematically
compared polarization maps and rotation of globules
\cite{Kane_1996}.
 They found that for 5 out of the 6 globules studied
the primary magnetic field direction was aligned with the
projected rotational axis, and the difference $|\theta_{\rm B} -
\theta_{\rm J}|$ was less than 20$^\circ$.
 These 5 globules included CB4 and CB17 with uniform
and well aligned magnetic fields nearly parallel to the Galactic
plane (all like in our case of CB67) and CB161, CB195, CB228 with
a bimodal distribution of polarization position angles.
 Kane and Clemens further suggest that some (primary)
polarization traces the field local to the globules, while other
polarization traces the field at some distance away from (most
likely behind) the globules.
 The secondary polarization usually appears to follow the
Galactic plane.
 The notable case is CB183 where two equally strong components
are observed one of which follows the Galactic plane while another
does not coincide with either magnetic field component.
 The authors conclude that CB183 may not be a simple
singly condensed globule, but a conglomeration of condensations.

The theoretical papers mainly included MHD calculations of
collapse for rotating spherical clouds when the magnetic fields
and the rotation axis are parallel and very seldom when they are
not \cite{Joo_2012}.
 In the latter case a disk is formed and
its angular rotation moment tends to be parallel either to the
magnetic fields  or to the rotation axis of the cloud depending on
the relative strength of the fields (see, for example, \cite{Mach_2006}).

The situation observed for CB67 when the cloud (core) is extended
and rotates around an axis which is almost perpendicular to the
magnetic field direction does not fit the theoretical
modeling.

\section{Conclusions}%
\label{sect5}

On the basis of polarimetric observations of about 50 stars
located in the vicinity of the Bok globule CB67 the orientation of
the magnetic fields near and in the outer layers of this cloud has
been considered.
 It is found that the globule and its core are extended and
rotate around the axis that makes the angle of about 70$^\circ$ with
the magnetic field direction.
 This situation is unusual for globules and can be hardly explained
in the framework of MHD calculations of collapse of a low mass
cloud.

\bigskip
\bigskip

The authors are thankful to an anonymous referee for
useful remarks which helped considerably to improve
the manuscript.

The authors thank Inter University of Center for Astronomy and
Astrophysics, Pune, India for allotment of telescope time for
doing this work.

This work was supported by the joint RFBR and DST (India) Grant
11-02-92695 (RUSP 1110) which supported an Russia--India
scientific collaboration, the RFBR Grant 13-02-00138 and a Grant
of the Minobrnayka of Russia within the state assignment for SUAI
in 2013--14.


\bibliography{prokopyeva}

\begin{thebibliography}{10}
\expandafter\ifx\csname url\endcsname\relax
  \def\url#1{\texttt{#1}}\fi
\expandafter\ifx\csname urlprefix\endcsname\relax\def\urlprefix{URL }\fi
\expandafter\ifx\csname href\endcsname\relax
  \def\href#1#2{#2} \def\path#1{#1}\fi

\bibitem{Ber_2007}
{Bergin E.A.}, {Tafalla M.}, Cold dark clouds: The initial conditions for star
  formation, Annual Review of Astronomy \& Astrophysics 45 (2007) 339--396.
\newblock \href {http://dx.doi.org/10.1146/annurev.astro.45.071206.100404}
  {\path{doi:10.1146/annurev.astro.45.071206.100404}}.

\bibitem{McK_2007}
{McKee C.F.}, {Ostriker E.C.}, Theory of star formation, Annual Review of
  Astronomy \& Astrophysics 45 (2007) 565--687.
\newblock \href {http://dx.doi.org/10.1146/annurev.astro.45.051806.110602}
  {\path{doi:10.1146/annurev.astro.45.051806.110602}}.

\bibitem{Cer_2011}
{Caselli P.}, Observational studies of pre-stellar cores and infrared dark
  clouds, in: {Cernicharo J.}, {Bachiller R.} (Eds.), The Molecular Universe,
  Vol. 280 of Proceedings of the International Astronomical Union, IAU
  Symposium, 2011, pp. 19--31.
\newblock \href {http://dx.doi.org/10.1017/S1743921311024835}
  {\path{doi:10.1017/S1743921311024835}}.

\bibitem{Hen_2013}
{Hennebelle P.}, Transport and influence of angular momentum in collapsing
  dense cores, in: {Hennebelle P.}, {Charbonnel C.} (Eds.), Angular Momentum
  Transport During the Formation and Early Evolution of Stars, Vol.~62 of EAS
  Publications Series, 2013, pp. 67--94.
\newblock \href {http://dx.doi.org/10.1051/eas/1362003}
  {\path{doi:10.1051/eas/1362003}}.

\bibitem{Cru_2012}
{Crutcher R.M.}, Magnetic fields in molecular clouds, Annual Review of
  Astronomy \& Astrophysics 50 (2012) 29--63.
\newblock \href {http://dx.doi.org/10.1146/annurev-astro-081811-125514}
  {\path{doi:10.1146/annurev-astro-081811-125514}}.

\bibitem{Ward_2009}
{Ward-Thompson D.}, {Sen A. K.}, {Kirk J. M.}, {Nutter D.}, Optical and
  submillimetre observations of bok globules - tracing the magnetic field from
  low to high density, Monthly Notices of the Royal Astronomical Society 398
  (2009) 394--400.
\newblock \href {http://dx.doi.org/10.1111/j.1365-2966.2009.15159.x}
  {\path{doi:10.1111/j.1365-2966.2009.15159.x}}.

\bibitem{Kane_1997}
{Kane, B. D.}, {Clemens, D. P}, Rotation of starless bok globules, Astronomical
  Journal 113 (1997) 1799--1814.
\newblock \href {http://dx.doi.org/10.1086/118392} {\path{doi:10.1086/118392}}.

\bibitem{Bel_2013}
{Belloche A.}, Observation of rotation in star forming regions: clouds, cores,
  disks, and jets, in: {Hennebelle P.}, {Charbonnel C.} (Eds.), Angular
  Momentum Transport During the Formation and Early Evolution of Stars, Vol.~62
  of EAS Publications Series, 2013, pp. 25--66.
\newblock \href {http://dx.doi.org/10.1051/eas/1362002}
  {\path{doi:10.1051/eas/1362002}}.

\bibitem{Kane_1996}
{Kane B.D.}, {Clemens D.}, Investigating correlations in the kinematics and
  magnetic fields of quiescent bok globules, in: {Roberge W.G.}, {Whittet
  D.C.B.} (Eds.), Polarimetry of Interstellar Medium, Vol.~97 of Astronomical
  Society of the Pacific Conference Series, 1996, p. 269.

\bibitem{Lom_2008}
{Lombardi M.}, {Lada C.J.}, {Alves J.}, Hipparcos distance estimates of the
  ophiuchus and the lupus cloud complexes, Astronomy \& Astrophysics 480 (2008)
  785--792.
\newblock \href {http://dx.doi.org/10.1051/0004-6361:20079110}
  {\path{doi:10.1051/0004-6361:20079110}}.

\bibitem{Cle_1988}
{Clemens D.P.}, {Barvainis R.}, A catalog of small, optically selected
  molecular clouds - optical, infrared, and millimeter properties,
  Astrophysical Journal Supplement Series 68 (1988) 257--286.
\newblock \href {http://dx.doi.org/10.1086/191288} {\path{doi:10.1086/191288}}.

\bibitem{Lyn_1962}
{Lynds B.T.}, Catalogue of dark nebulae, Astrophysical Journal Supplement
  Series 7 (1962) 1.
\newblock \href {http://dx.doi.org/10.1086/190072} {\path{doi:10.1086/190072}}.

\bibitem{Noz_1991}
{Nozawa S.}, {Mizuno A.}, {Teshima Y.}, {Ogawa H.}, {Fukui Y.}, A study of
  (c-13)o cloud cores in ophiuchus, Astrophysical Journal Supplement Series 77
  (1991) 647--675.
\newblock \href {http://dx.doi.org/10.1086/191618} {\path{doi:10.1086/191618}}.

\bibitem{Lau_1997}
{Launhardt R.}, {Henning T.}, Millimetre dust emission from northern bok
  globules, Astronomy \& Astrophysics 326 (1997) 329--346.

\bibitem{Vis_2001}
{Visser A. E.}, {Richer J. S.}, {Chandler C. J.}, A scuba survey of compact
  dark lynds clouds, Monthly Notices of the Royal Astronomical Society 323
  (2001) 257--269.
\newblock \href {http://dx.doi.org/10.1046/j.1365-8711.2001.04037.x}
  {\path{doi:10.1046/j.1365-8711.2001.04037.x}}.

\bibitem{Arz_2011}
{Arzoumanian D. et al.}, Characterizing interstellar filaments with herschel in
  ic 5146, Astronomy \& Astrophysics 529 (2011) L6.
\newblock \href {http://dx.doi.org/10.1051/0004-6361/201116596}
  {\path{doi:10.1051/0004-6361/201116596}}.

\bibitem{Han_2013}
{Han J.}, Magnetic fields in our milky way galaxy and nearby galaxies, in:
  Solar and Astrophysical Dynamos and Magnetic Activity, Vol. 294 of
  Proceedings of the International Astronomical Union, IAU Symposium, 2013, pp.
  213--224.
\newblock \href {http://dx.doi.org/10.1017/S1743921313002561}
  {\path{doi:10.1017/S1743921313002561}}.

\bibitem{Pavel_2012}
{Pavel M. D.}, {Clemens D. P.}, {Pinnick A. F.}, Testing galactic magnetic
  field models using near-infrared polarimetry, The Astrophysical Journal 749
  (2012) 71.
\newblock \href {http://dx.doi.org/10.1088/0004-637X/749/1/71}
  {\path{doi:10.1088/0004-637X/749/1/71}}.

\bibitem{Heiles_2000}
{Heiles C.}, 9286 stars: An agglomeration of stellar polarization catalogs, The
  Astronomical Journal 119 (2000) 923--927.
\newblock \href {http://dx.doi.org/10.1086/301236} {\path{doi:10.1086/301236}}.

\bibitem{Choi_2013}
{Choi Y.-J.}, {Min K.-W.}, {Seon K.-I.}, {Lim T.-H.}, {Jo Y.-S.}, {Park J.-W.},
  Far-ultraviolet observations of the spica nebula and the interaction zone,
  The Astrophysical Journal 774 (2013) 34.
\newblock \href {http://dx.doi.org/10.1088/0004-637X/774/1/34}
  {\path{doi:10.1088/0004-637X/774/1/34}}.

\bibitem{Santos_2011}
{Santos F. P.}, {Corradi W.}, {Reis W.}, Optical polarization mapping toward
  the interface between the local cavity and loop i, The Astrophysical Journal
  728 (2011) 104.
\newblock \href {http://dx.doi.org/10.1088/0004-637X/728/2/104}
  {\path{doi:10.1088/0004-637X/728/2/104}}.

\bibitem{Li_2013}
{Li H.}, {Fang M.}, {Henning T.}, {Kainulainen J.}, The link between magnetic
  fields and filamentary clouds: bimodal cloud orientations in the gould belt,
  Monthly Notices of the Royal Astronomical Society 436 (2013) 3707--3719.
\newblock \href {http://dx.doi.org/10.1093/mnras/stt1849}
  {\path{doi:10.1093/mnras/stt1849}}.

\bibitem{Sen_1994}
{Sen A. K.}, {Tandon S. N.}, Two-channel optical imaging polarimeter, in:
  {David L. C.}, {Eric R. C.} (Eds.), Instrumentation in Astronomy VIII, Vol.
  2198 of Proc. SPIE, 1994, pp. 264--273.

\bibitem{Ramaprakash_1998}
{Ramaprakash A. N.}, {Gupta R.}, {Sen A. K.}, {Tandon S. N.}, An imaging
  polarimeter (impol) for multi-wavelength observations, Astronomy and
  Astrophysics Supplement 128 (1998) 369--375.
\newblock \href {http://dx.doi.org/10.1051/aas:1998148}
  {\path{doi:10.1051/aas:1998148}}.

\bibitem{Pau_2012}
{Paul D.}, {Das H. S.}, {Sen, A. K.}, Imaging polarimetry of the bok globule
  cb56, Bulletin of the Astronomical Society of India 40 (2012) 113--119.

\bibitem{Tur_1990}
{Turnshek D.A. et al.}, An atlas of hubble space telescope photometric,
  spectrophotometric, and polarimetric calibration objects, The Astronomical
  Journal 99 (1990) 1243--1261.
\newblock \href {http://dx.doi.org/10.1086/115413} {\path{doi:10.1086/115413}}.

\bibitem{Hsu_1982}
{Hsu J.}, {Breger M.}, On standard polarized stars, The Astrophysical Journal
  262 (1982) 732--738.
\newblock \href {http://dx.doi.org/10.1086/160467} {\path{doi:10.1086/160467}}.

\bibitem{Eswaraiah_2013}
{Eswaraiah C.}, {Maheswar G.}, {Pandey A. K.}, {Jose J.}, {Ramaprakash A. N.},
  {Bhatt H. C.}, A study of the starless dark cloud ldn 1570: Distance, dust
  properties, and magnetic field geometry, Astronomy and Astrophysics 556
  (2013) A65.
\newblock \href {http://dx.doi.org/10.1051/0004-6361/201220603}
  {\path{doi:10.1051/0004-6361/201220603}}.

\bibitem{Skrutskie_2006}
{Skrutskie, M. F. et. all}, The two micron all sky survey (2mass), The
  Astronomical Journal 131 (2006) 1163--1183.
\newblock \href {http://dx.doi.org/10.1086/498708} {\path{doi:10.1086/498708}}.

\bibitem{Maheswar_2010}
{Maheswar G.}, {Lee C. W.}, {Bhatt H. C.}, {Mallik S. V.}, {Dib S.}, A method
  to determine distances to molecular clouds using near-ir photometry,
  Astronomy and Astrophysics 509 (2010) A44.
\newblock \href {http://dx.doi.org/10.1051/0004-6361/200912594}
  {\path{doi:10.1051/0004-6361/200912594}}.

\bibitem{Rowles_2009}
{Rowles J.}, {Froebrich D.}, The structure of molecular clouds - i. all-sky
  near-infrared extinction maps, Monthly Notices of the Royal Astronomical
  Society 395 (2009) 1640--1648.
\newblock \href {http://dx.doi.org/10.1111/j.1365-2966.2009.14655.x}
  {\path{doi:10.1111/j.1365-2966.2009.14655.x}}.

\bibitem{McCutcheon_1986}
{McCutcheon W. H.}, {Vrba F. J.}, {Dickman R. L.}, {Clemens D. P.}, The lynds
  204 complex - magnetic field controlled evolution?, Astrophysical Journal 309
  (1986) 619--627.
\newblock \href {http://dx.doi.org/10.1086/164630} {\path{doi:10.1086/164630}}.

\bibitem{Vergely_2010}
{Vergely J.-L.}, {Valette B.}, {Lallement R.}, {Raimond S.}, Spatial
  distribution of interstellar dust in the sun's vicinity. comparison with
  neutral sodium-bearing gas, Astronomy and Astrophysics 518 (2010) A31.
\newblock \href {http://dx.doi.org/10.1051/0004-6361/200913962}
  {\path{doi:10.1051/0004-6361/200913962}}.

\bibitem{Welsh_2010}
{Welsh B. Y.}, {Lallement R.}, {Vergely J.-L.}, {Raimond S.}, New 3d gas
  density maps of nai and caii interstellar absorption within 300 pc, Astronomy
  and Astrophysics 510 (2010) A54.
\newblock \href {http://dx.doi.org/10.1051/0004-6361/200913202}
  {\path{doi:10.1051/0004-6361/200913202}}.

\bibitem{Hilton_1995}
{Hilton J.}, {Lahulla J. F.}, Distance measurements of lynds galactic dark
  nebulae., Astronomy and Astrophysics Supplement 113 (1995) 325.

\bibitem{Joo_2012}
{Joos M.}, {Hennebelle P.}, {Ciardi A.}, Protostellar disk formation and
  transport of angular momentum during magnetized core collapse, Astronomy \&
  Astrophysic 543 (2012) A128.
\newblock \href {http://dx.doi.org/10.1051/0004-6361/201118730}
  {\path{doi:10.1051/0004-6361/201118730}}.

\bibitem{Mach_2006}
{Machida M. N.}, {Matsumoto T.}, {Hanawa T.}, {Tomisaka K.}, Evolution of
  rotating molecular cloud core with oblique magnetic field, The Astrophysical
  Journal 645 (2006) 1227--1245.
\newblock \href {http://dx.doi.org/10.1086/504423} {\path{doi:10.1086/504423}}.

\end{thebibliography}

\end{document}